\documentclass[%
 reprint,
 groupedaddress
 amsmath,amssymb,
 aps,
 prl,
]{revtex4-1}

\usepackage{graphicx}
\usepackage{dcolumn}
\usepackage{bm}
\usepackage{here}
\usepackage{color}
\usepackage{ulem}

\begin{document}

\preprint{APS/123-QED}

\title{Successive field-induced transitions in BiFeO$_{3}$ around room temperature}

\author{Shiro Kawachi,$^1$ Atsushi Miyake,$^1$ Toshimitsu Ito,$^2$ Sachith E. Dissanayake,$^3$ Masaaki Matsuda,$^3$ William Ratcliff II,$^4$ Zhijun Xu,$^{4,5}$ Yang Zhao,$^{4,5}$ Shin Miyahara,$^6$ Nobuo Furukawa,$^7$ and Masashi Tokunaga$^1$}

\affiliation{\vspace{4pt}$^1$The Institute for Solid State Physics (ISSP), The University of Tokyo, Kashiwa, Chiba 277-8581, Japan.
\\$^2$National Institute of Advanced Industrial Science and Technology (AIST), Tsukuba, Ibaraki 305-8562, Japan.
\\$^3$Quantum Condensed Matter Division, Oak Ridge National Laboratory (ORNL), Oak Ridge, Tennessee 37831, USA.
\\$^4$NIST center for Neutron Research, NIST, Gaithersburg, Maryland 20899, USA.
\\$^5${Department of Materials Science and Engineering, University of Maryland, College Park, Maryland 20742, USA.}
\\$^6${Department of Applied Physics, Fukuoka University, Jonan-ku, Fukuoka 814-0180, Japan.}
\\$^7$Department of Physics and Mathematics, Aoyama Gakuin University, Sagamihara, Kanagawa 229-8558, Japan.}

\begin{abstract}

The effects of high magnetic fields applied perpendicular to the spontaneous ferroelectric polarization on single crystals of BiFeO$_3$ were investigated through magnetization, magnetostriction, and neutron diffraction measurements. The magnetostriction measurements revealed lattice distortion of $2\times 10^{-5}$, during the reorientation process of the cycloidal spin order by applied magnetic fields. Furthermore, anomalous changes in magnetostriction and electric polarization at a larger field demonstrate an intermediate phase between cycloidal and canted antiferromagnetic states, where a large magnetoelectric effect was observed. Neutron diffraction measurements clarified that incommensurate spin modulation along [110] direction in the cycloidal phase becomes commensurate in the intermediate phase. Theoretical calculations based on the standard spin Hamiltonian of this material suggest an antiferromagnetic cone-type spin order in the intermediate phase.
 
\end{abstract}

\maketitle


Recently, multiferroic materials have been widely investigated due to their coupling between magnetic and ferroelectric ordering.  BiFeO$_3$ is perhaps the most extensively studied multiferroic material as it possesses robust multiferroicity at room temperature as well as various possible applications \cite{kezmarki2015,chu2008,choi2009,wu2010,jiang2011,wang2011,tsurumaki2012,hong2013,kawachi2016,yang2009}.
The effects of magnetic fields on the coupled multiple degrees of freedom behind these phenomena are not fully understood.

\par BiFeO$_3$ exhibits a cycloidal magnetic order below 640 K \cite{sosnovska1982}. This state is known to exhibit marginal quadratic magnetoelectric (ME) effect at low magnetic fields \cite{munoz1985}.
High magnetic fields of $\sim$20 T stabilize the canted antiferromagnetic (CAFM) phase as opposed to the cycloidal phase \cite{popov1993, kadomtseva1995, ruette2004, tokunaga2015jmmm, tokunaga2015}. 
Although several groups succeeded in realizing the CAFM phase at zero field \cite{chen2012,Sosnowska2013,yamamoto2016,bai2005,bea2005,Sando2013}, the ME effect in this phase has not been clarified.
In this study, several features were observed when a magnetic field was applied normal to the trigonal $c$-axis, and they indicated that a third magnetic phase emerged in bulk BiFeO$_3$ between the cycloidal and CAFM phases at approximately room temperature.


\par BiFeO$_3$ has a crystal structure with the polar space group of $R3c$.
A large switchable spontaneous electric polarization emerges along the $c$-axis of the trigonal cell \cite{Shvartsman2007, Lebeugle2007, Catalan2009}. Degeneracy exists in selecting the polarization direction from eight $\langle111\rangle$ directions in the cubic unit. Hence, depending on synthesis methods, BiFeO$_3$ crystals can contain multiple ferroelectric domains \cite{slee2008prb}.

\par Magnetic domains can be present even in single ferroelectric domain crystals. The magnetic propagation vector $Q$ points in one of the $\langle110\rangle$ directions of the trigonal cell \cite{sosnovska1982} in the cycloidal spin ordered state below $\sim$640 K. The three-fold rotational symmetry around the $c$-axis ($Z$ direction in this paper) leads to three equivalent $Q_i$ $(i = 1,2,3)$, as shown in Fig. 1(d). The spins rotated primarily in the $Q_i$-$Z$ plane in a magnetic domain with a given $Q_i$.


\par Recently, Tokunaga $et$ $al$. indicated the emergence of an electric polarization perpendicular to the $Z$ direction that was controlled by magnetic fields \cite{tokunaga2015}. Theoretical calculations suggested that the cycloidal spin order in BiFeO$_3$ could involve electric polarization perpendicular to the $Q_i$-$Z$ plane as illustrated as $P\rm_T$ in Fig. 1(d) \cite{lee2015,miyahara2016,kaplan2011}. The existence of $P\rm_T$ indicated that three-fold rotational symmetry was broken in the cycloidal state. Therefore, BiFeO$_3$ had lower symmetry than $R3c$ at room temperature. Sosnowska $et$ $al$. examined the crystal structure of BiFeO$_3$ using synchrotron X-ray diffraction and proposed that monoclinic distortion led to the observed broadening of the Bragg peaks below 1038 K \cite{sosnowska2012}. However, relation between the magnetic order and the monoclinic distortion was not clear since this broadening was observed even at temperatures well above 640 K.


\par In this study, the magnetization and magnetostriction of single ferroelectric domain crystals of BiFeO$_3$ synthesized by the laser-diode heating floating-zone method \cite{Ito2011} were measured in pulsed high magnetic fields at ISSP. The magnetization were measured by the induction method. Newly improved capacitance dilatometry enabled the measurement of magnetostriction using the capacitance method \cite{kido1989}. The neutron diffraction experiments were carried out on the BT-7 thermal neutron triple-axis spectrometer at the NIST Center for Neutron Research \cite{bt7}. 
The magnetic domains of the crystal were previously aligned by applying magnetic field along the $Y$ direction [see Fig. 1(d)]. The single crystal was oriented 
\begin{flushleft}
\begin{figure}[H]
 \centering
 \includegraphics[width=8.6cm]{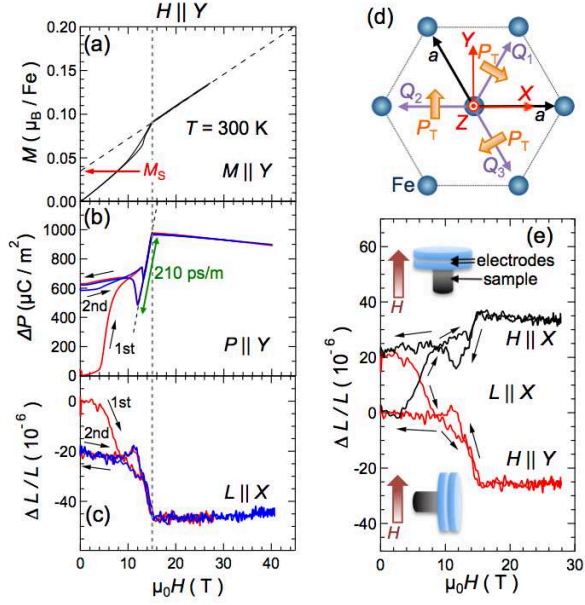} 
 \caption{Magnetic field-induced changes in (a) magnetization, (b) electric polarization along the $Y$ direction, and (c) magnetostriction along the $X$ direction at 300 K for $H\parallel Y$. In (b) and (c), virgin and second traces are presented by red and blue lines, respectively. (d) Schematic drawing of the arrangement of Fe ions, coordinates, magnetic $Q-$vectors, and transverse electric polarization ($P\rm _T$) in the $ab$-plane of the trigonal cell. (e) Magnetostriction along $X$ direction in magnetic fields parallel to $X$ and $Y$ at 300 K. 
 }\label{fig:fig1}
\end{figure}
\end{flushleft}
in the ($h{h}l$) $\lbrack$or ($X0Z$)$\rbrack$ scattering plane and was mounted in a 15 T vertical field superconducting magnet. The magnetic field was applied along $Y$ direction.
Details of the neutron experiment are described in the supplementary material \cite{supplementary}.


\par Figure 1(a) shows the magnetization ($M$) curve at 300 K in the magnetic field ($H$) applied along the $Y$ direction. The kink at $\sim$ 15 T indicated the existence of a magnetic transition at this field \cite{tokunaga2015}. Extrapolation of the linear $M$-$H$ curve at high field to zero field indicated finite offsets in the vertical axis and thus suggested that the high field phase possesses spontaneous magnetization ($M_S$). Figure 1(b) presents the change in electric polarization along the $Y$ direction ($\Delta P_Y$) caused by a magnetic field $H\parallel Y$. The $\Delta P_Y$ steeply changes below 10 T in the field increasing process of the first field cycle (red). However, this change was suppressed in the second field cycle (blue). This irreversible change (also known as the non-volatile effect) was ascribed to the reorientation of the magnetoelectric domains \cite{kawachi2016, tokunaga2015}. It was assumed that 
the transverse electric polarization $P_{\rm T}$ changed the direction during the reorientation of the cycloidal domains to the $Q_2$ domain in the initial field scan.


\par The emergence of $P_Y$ indicates that three-fold 
\begin{flushleft}
\begin{figure}[t]
 \centering
 \includegraphics[width=8.6cm]{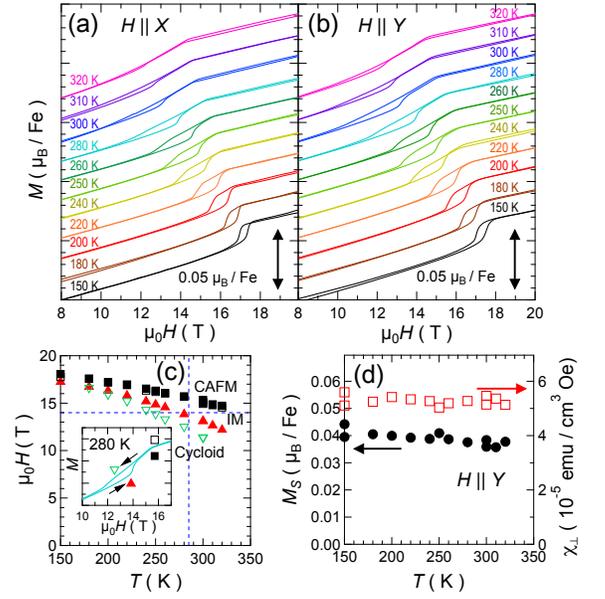} 
 \caption{Magnetization curves at various temperatures in magnetic fields applied along (a) $X$ and (b) $Y$ direction. Data obtained at different temperatures are vertically offset for clarity. (c) Temperature dependence of transition fields for $H\parallel$ $Y$. The transition fields are defined in the $M$-$H$ curves marked by symbols in the inset. Open and closed symbols correspond to the transition fields of the field increasing and decreasing processes, respectively. The
blue dashed lines indicate the highest field (14 T) and temperature (285 K) for the neutron experiments in Figs. 4. (d) Temperature dependence of the spontaneous magnetization ($M_{\rm S}$, closed circles) and differential magnetic susceptibility ($\chi_\perp$, open squares) in the CAFM phase.}\label{fig:fig2}
\end{figure}
\end{flushleft}
rotational symmetry was broken at 300 K. We measured the transverse magnetostriction with $H\parallel $ $Y$ at 300 K to detect the relevant lattice distortion. The vertical axis in Fig. 1(c) corresponds to the field-induced change in the sample length along $X$ ($\Delta L$) normalized by the length at zero field ($L$). The field dependence of $\Delta L/L$ followed that of $\Delta P_Y$, in the opposite sign, including the irreversible behavior below 10 T. This measurement demonstrated that the application of a temporal magnetic field slightly compressed the crystal normal to the $H$-$Z$ plane. Prior to this experiment, we measured the magnetostriction along the $X$ direction for $H\parallel$ $X$ while maintaining the capacitance cell set-up [insets of Fig. 1(e)]. This result is indicated by a black solid line in Fig. 1(e). In the case of $H\parallel$ $X$, the crystal length along the $X$ direction increased. The initial position of the $\Delta L/L$ for $H\parallel Y$ was the same as its final position after the application of $H\parallel X$. Therefore, the data for $H\parallel Y$ is plotted with the vertical offset of $2\times10^{-5}$. These irreversible behaviors below 10 T indicated that the sample length changed due 
\begin{flushleft}
\begin{figure}[H]
 \centering
 \includegraphics[width=8.6cm]{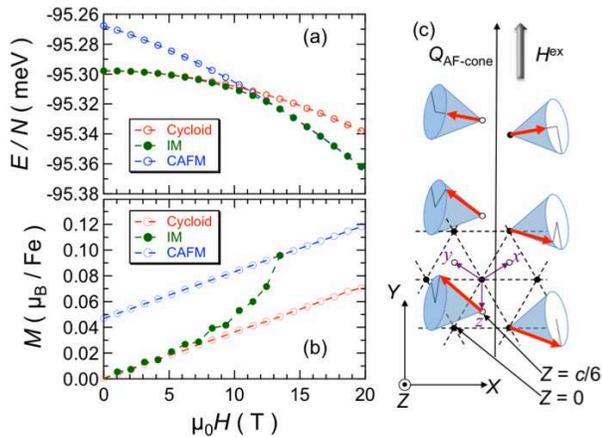} 
 \caption{Calculated magnetic field dependences of (a) energy per site and (b) magnetization in the cycloidal, intermediate (IM), and CAFM phases. \color{black}(c) Schematic drawing of the AF-cone spin structure expected in the IM phase. Closed and open circles represent Fe sites which differ by ($c/6$) along the $Z$-axis. Red arrows represent the spin moments. $x$, $y$, and $z$ denote three adjacent directions in the pseudocubic unit. The propagation vector $Q_{\rm AF-cone}$ becomes parallel to the field $H^{\rm ex}$ applied in the $Y$ direction.}\label{fig:fig3}
\end{figure}
\end{flushleft}
to the direction of $\vec{Q}$. A neutron experiment in a recent study confirmed that application of $H \parallel Y$ of 6 T would stabilize the $Q_2$ domain in Fig. 1(d) \cite{matsuda_unpub}. Contraction of the sample along the $X$ direction for $H \parallel Y$ indicated that the sample slightly contracted along the $Q$ vector. 
This result is consistent with the early report of neutron diffraction measurements showing that the population of the magnetic domain with the $Q$ vector normal to the applied pressure direction decreased when uniaxial pressure was applied along the principal axis of the pseudocubic unit cell \cite{Ramazanoglu2011}. 


\par A systematic analysis of the data obtained after the domain reorientation in Figs. 1(a)-(c) indicated that all the quantities showed non-monotonic changes between 10 T and 15 T. In order to clarify this anomaly, we measured the magnetization with $H\parallel X$ and $H\parallel Y$ at various temperatures.  The results are shown in Figs. 2(a) and 2(b). 
As observed in these figures, no intrinsic differences were detected in the $M$-$H$ curves for $H\parallel X$ and $H\parallel Y$. 
At 150 K, the magnetization showed a step-like change at the transition field. Conversely, $M$-$H$ curves above 200 K showed the region with a large gradient in the intermediate field region.  In the following, we will refer to the magnetic phase in this intermediate region as the IM phase. The transition from the cycloidal phase to the IM phase was discontinuous while that from the IM phase to the CAFM phase was continuous. The IM phase appeared from the lower field as the temperature increased. Figure 2(c) shows the phase diagram in the $H$-$T$ plane determined by the magnetization curves. Here, characteristic fields were assigned in the magnetization curves as the transition fields as marked by several symbols in the inset of Fig. 2(c). 


\par Gareeva $et$ $al$. theoretically predicted the emergence of the intermediate phase \cite{gareeva2013}. Their study proposed that an antiferromagnetic cone (AF-cone) phase existed between the cycloidal and CAFM phases in the magnetic fields parallel as well as perpendicular to the trigonal axis. The calculations suggested that the boundary between the cycloidal and AF-cone phases corresponded to a first order phase transition, and this is consistent with the results of the present study. Their calculations focused on a one-dimensional spatially modulated spin structure along a fixed direction. Therefore, it is necessary to consider more general states in order to determine the actual spin order in the IM phase. 


\par We performed calculations starting with the following spin Hamiltonian \cite{Jeong2012,Matsuda2012,Fishman2012}:
\begin{eqnarray*}\label{MiyaharaHami}
\mathcal{H} &=&J_1 \sum_{\rm n.n.}{\bm S}_i \cdot {\bm S}_j + J_2 \sum_{\rm n.n.n}{\bm S}_i \cdot {\bm S}_j + \sum_{\rm n.n.} \biggl\{ -\frac{1}{2}D\big(S_i^Y S_{i+x}^Z \nonumber\\
& & - S_i^Z S_{i+x}^Y \big) + \frac{\sqrt{3}}{2}D\left(S_i^Z S_{i+x}^X - S_i^X S_{i+x}^Z \right) \nonumber \\
& &-\frac{1}{2} D \left(S_i^Y S_{i+y}^Z - S_i^Z S_{i+y}^Y \right) -\frac{\sqrt{3}}{2} D \big(S_i^Z S_{i+y}^X \nonumber \\
& &- S_i^X S_{i+y}^Z \big) + D \left(S_i^Y S_{i+z}^Z - S_i^Z S_{i+z}^Y \right)\biggr\}\nonumber \\
& &+ \sum_{\rm n.n}\left(-1\right)^{n_i} D^{\prime} \left(S_i^X S_j^Y - S_i^Y S_j^X\right)\nonumber \\
& &-g\mu_B \sum {\bm B} \cdot {\bm S}_i - K \sum\left(S_i^Z\right)^2.
\end{eqnarray*}
Here, $i$, $j$ and $x$, $y$, $z$ represent the Fe sites and three adjacent directions in the pseudocubic cell as shown in Fig. 3(c), respectively. $J_1$ and $J_2$ represent nearest and next-nearest neighbor interaction. $D$, $D^{\prime}$, and $K$ denote DM interactions and single-ion anisotropy. 
Figure 3(a) shows the energy of several phases as a function of magnetic field applied normal to the $Z$ direction with typical values of parameters \cite{parameters}. An intermediate (IM) phase is stabilized between the cycloidal and CAFM phases.
The IM phase shows an AF cone-type spin structure with the propagation vector pointing to the field direction as shown in Fig. 3(c). In the present Hamiltonian, the $D$ terms stabilize the cycloidal state, while the $D'$ term favors the CAFM one. The AF cone state is a kind of the superposition of these two orders caused by the competition of these two DM interactions. Figure 3(b) shows calculated magnetization curves for the three phases. The gradual increase in magnetization for the IM phase reproduces the experimentally observed change in the slope shown in Fig. 2(b).


\par In this AF-cone state, spin susceptibility along the spin modulation vector will be larger than that normal to it. Accordingly, a flop of the modulation vector can be 
\begin{flushleft}
\begin{figure}[H]
 \centering
 \includegraphics[width=8.6cm]{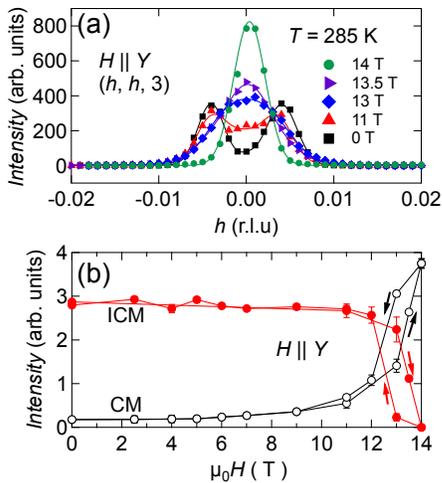} 
 \caption{(a) Profiles of the neutron diffraction along the $X$ direction in applied magnetic field along the $Y$ direction at 285 K, measured upon increasing the magnetic field. Symbols and solid lines are the experimental data and the fits to the data, respectively. The errorbars are smaller than the symbol size. (b) Relative integrated intensities of the commensurate (open circles) and  the incommensurate peaks (solid circles) as a function of the applied field. Solid lines are the guide to the eyes.}\label{fig:fig4}
\end{figure}
\end{flushleft}
caused by the transition from the cycloidal to the AF-cone state.
Therefore, we performed neutron diffraction measurements of BiFeO$_3$ to study the field dependence of the spin modulation through this transition.
We performed scans along the $X$ direction at 285 K under applied $H\parallel Y$. The data were taken as we ramped up the magnetic field and these scans are shown in Figure 4(a).   As the field increases, the incommensurate peaks at $h$ $\sim \pm$0.004 are reduced in intensity, while in contrast, the commensurate peak at $h$ = 0 grows. In Fig. 4(b), we show the field dependence of the integrated intensities of the commensurate and incommensurate peaks. At 285 K, application of 14 T was sufficient to reach the IM phase, but insufficient to reach the CAFM phase, at this temperature as denoted by a cross point of vertical and horizontal dashed lines in Fig. 2(c).
An increase in magnetic fields from $\sim$6 T led to a gradual increase and decrease in the commensurate and incommensurate scattering intensities, respectively. The spin structure appeared commensurate along the $X$ direction in the IM phase realized at 14 T. By using the results of the present study, it is not possible to conclude whether the spin structure was commensurate or incommensurate along the $Y$ direction due to the limited resolution out of the scattering plane. A decrease in the magnetic field resulted in the change from the commensurate state to the incommensurate state showing hysteresis. The coexistence of commensurate/incommensurate scattering supports the conclusion that the transition from the cycloidal phase to the IM phase is first order. The disappearance of the incommensurate spin modulation along the $X$ direction in the IM phase did not contradict the picture of the AF-cone phase as shown in Fig. 3(c). We also performed neutron diffraction measurements with $H \parallel X$ and found that the transition to the IM phase occurred at $\sim$14 T \cite{supplementary}, consistent with the magnetization results.


\par In the IM phase, magnetization, electric polarization, and magnetostriction changed linearly as a function of magnetic field. The coefficient of the linear ME effect between 12 T and 15 T shown in Fig. 1(b) is approximately 210~ps/m. 
Although we cannot identify this value determined in a limited field region as the representative one in the IM phase, the observed large ME effect implies the presence of strong ME coupling in this phase.


\par Thus far, we have not succeeded in reproducing the ME effect in the IM phase based on the generalized inverse Dzyaloshinskii-Moriya effect \cite{miyahara2016}. Phenomenologically, the ME effect might be explained by slight tilting of the trigonal $c$-axis by the monoclinic distortion. Using the reported lattice parameters \cite{sosnowska2012}, we determined that the tilting angle was $\theta=0.01^\circ-0.04^\circ$ \cite{supplementary}. Assuming a spontaneous polarization of 1C$\!/$m$^2$ along the $c$-axis, the projected component will be between 200 and 800 $\rm\mu$C$\!/$m$^2$, which is the same order of magnitude with the observed transverse component. The important point here is that the tilting is coupled to the spin system, and hence, can be controlled by an external magnetic field.


\par Finally, the study involved examining ways to realize the IM phase at lower magnetic fields than the fields observed in this study. In accordance with the Landau-Ginzburg theory for BiFeO$_3$, exchange stiffness, DM interaction, magnetic anisotropy constant ($\kappa_c$), differential magnetic susceptibility ($\chi_\perp$), and spontaneous magnetization in the CAFM phase ($M_{\rm S}$) are fundamental parameters. Among these parameters, $\chi_\perp$ and $M_{\rm S}$ were determined from the experimental results of the $M$-$H$ curves. Figure 2(d) shows the temperature dependence of these parameters determined from the magnetization curves at various temperatures. These parameters were almost constant at temperatures from 150 K to 320 K. The temperature dependence of the transition field to the IM phase could be a result of the change in $\kappa_c$ because the exchange stiffness and DM interaction are likely insensitive to temperature. Such temperature dependence of $\kappa_c$ may appear as the change in anharmonicity in the cycloidal order, which is observed experimentally \cite{Ramazanoglu2011b}. According to a theoretical study \cite{gareeva2013}, when $\kappa_c$ was increased, while holding the other parameters constant, the transition field to the AF-cone phase was reduced. Therefore, future research could examine the AF-cone phase in thin films with large $\kappa_c$. 


\par In summary, magnetization, magnetostriction, and neutron diffraction measurements were performed on single BiFeO$_3$ crystals in magnetic fields applied normal to the trigonal axis. The observations indicated that the reorientation of multiferroic domains occurred below 10 T in conjunction with irreversible field-induced change in the lattice distortion. This demonstrated the existence of a monoclinic distortion in the cycloidal phase. The application of an increased magnetic field realized an intermediate phase prior to the transition to the canted antiferromagnetic state. A large ME effect was observed in this intermediate phase. Theoretical calculations indicated the presence of an antiferromagnetic cone-type spin order in this phase, consistent with the neutron diffraction measurements.


\par This work was supported by the MEXT of Japan Grant-in-Aid for Challenging Exploratory Research (16K05413), and Murata Science Foundation.

\end{document}